\documentclass[twocolumn]{aa}
\usepackage{graphicx}
\usepackage{txfonts}

\def\ltsima{$\; \buildrel < \over \sim \;$}
\def\simlt{\lower.5ex\hbox{\ltsima}}
\def\gtsima{$\; \buildrel > \over \sim \;$}
\def\simgt{\lower.5ex\hbox{\gtsima}}
\begin{document}
   \title{Detection of a population gradient in the Sagittarius Stream}

   \author{M. Bellazzini\inst{1}, H.J. Newberg\inst{2}, M. Correnti\inst{1}, 
           F.R. Ferraro\inst{3}\and L. Monaco\inst{4}
          \fnmsep\thanks{Based on observations collected at the European Southern 
	  Observatory, Chile, (Programme 71.D-0222A).}
          }

   \offprints{M. Bellazzini}

   \institute{INAF - Osservatorio Astronomico di Bologna,
              Via Ranzani 1, 20127, Bologna, ITALY \\
              \email{michele.bellazzini@oabo.inaf.it}
         \and
	     Department of Physics and Astronomy, Rensselaer Polytechnic 
	     Institute, Troy, NY 12180;
         \and
             Dipartimento di Astronomia, Universit\`a di Bologna,
	     Via Ranzani 1, 20127, Bologna, ITALY 
         \and
	     ESO - European Southern Observatory, Alonso de Cordova 3107, 
	     Santiago 19, CHILE
             }

   \date{Received July 10, 2006; accepted August 16, 2006}

   \abstract{We present a quantitative comparison between the Horizontal Branch
   morphology in the core of the Sagittarius dwarf spheroidal galaxy (Sgr) and
   in a wide field  sampling a portion of its tidal stream (Sgr Stream), located
   tens of kpc away from the center of the parent galaxy. We find that the
   Blue Horizontal Branch (BHB)  stars in that part of the Stream are five
   times more abundant than in the Sgr core, relative to Red Clump stars.
   The difference in the ratio of BHB to RC stars between the two fields 
   is significant at the $\simgt 4.8\sigma$ level.
   This indicates that the old and metal-poor  population of Sgr was
   preferentially stripped from the galaxy in past peri-Galactic passages with
   respect to the intermediate-age metal rich population that presently dominates the
   bound core of Sgr, probably due to a strong radial gradient that was settled
   within the galaxy before its disruption.  
   The technique adopted in the present study allows to trace population
   gradients along the whole extension of the Stream. 

   }
    
   \authorrunning{M. Bellazzini et al.}
   \titlerunning{A population gradient in the Sagittarius Stream} 
    
   \maketitle
%

\section{Introduction}

The Sagittarius dwarf spheroidal galaxy (Sgr dSph; Ibata et al. \cite{iba1})
provides an excellent case for the study of the process of tidal disruption and
accretion of a dwarf satellite into a large galaxy. Its huge tidal tails form a
coherent and  dynamically cold filamentary structure (hereafter Sgr Stream)
extending for tens of kpc from the parent galaxy that has been probed with 
many different tracers (see, among the others, Ibata  et al. \cite{iba2}, 
Newberg et al. \cite{sdss} - hereafter N02, Majewski et al. \cite{maj03} - 
hereafter M03, Mart\'inez-Delgado et al. \cite{david} - hereafter MD04, 
Vivas et al. \cite{vivas}, Belokurov et al. \cite{belok}, and references 
therein).


%
   \begin{figure}
   \centering
   \includegraphics[width=8.5cm]{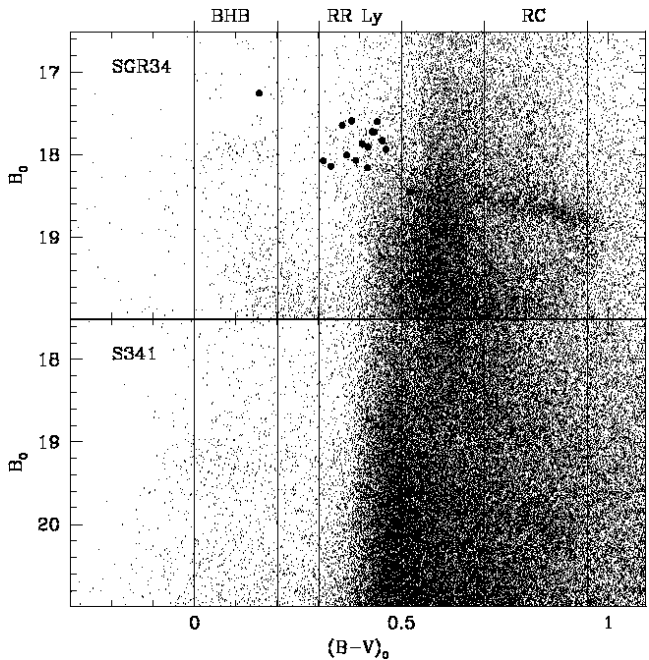}
      \caption{CMDs focused on the Horizontal Branch for fields in the core of 
      Sgr
   (upper panel) and in the Sgr Stream (lower panel). The color strips adopted
   to select candidate RC and BHB stars are enclosed within vertical lines and
   labeled. The filled circles in the CMD of SGR34 are the ab type RR Lyrae
   counter-identified in the nearby SGRW field from the catalogue of Cseresnjes 
   (\cite{cser}).
              }
         \label{HBSEL}
   \end{figure}
%


It is quite clear that the progenitor of the presently observed Sgr remnant
was a quite remarkable galaxy, possibly similar to the LMC (Monaco
et al. 2005, hereafter M05).  While the stellar content of the bound core of
the galaxy is dominated by an  intermediate-old metal-rich population
($[M/H]\sim -0.5$, M05, Bellazzini et al. \cite{conun}), 
the globular clusters possibly
associated with the Stream are predominantly old and metal-poor (Bellazzini,
Ferraro \& Ibata \cite{clus}; see also Belokurov et al. \cite{belok}). 
M03 and MD04 provided the first indications of the possible presence of a
metallicity gradient along the Stream, suggesting that the stars lost in
previous peri-Galactic passages, were, on average, more metal poor than those
recently lost or still bound to the main body. Chou et al.
(\cite{chou} - hereafter C06 ), studying the chemical abundances of relatively 
nearby Stream
stars, have shown that the metallicity distribution within the Stream is
significantly different from that of the main body of Sgr, the former
being skewed toward lower metallicities. These results strongly
suggest that the stellar content of the Sgr progenitor could have been quite
different from the present-day remnant, and indicates that it may be possible
to trace the change of stellar content along the Stream (see C06).  

The Horizontal Branch (HB) morphology is a very powerful tool to study
population gradients within galaxies (see Harbeck et al. 2001), since the color
of HB stars depends very strongly on their age and metal content 
(see Fusi Pecci et al. \cite{ffp1} for discussion and references). 
In particular, HB
stars can be located to the {\em blue} of the RR Lyrae instability strip (Blue
HB stars) only  if they are very old (age $>$ 10 Gyr) and (typically) 
metal-poor, while HB
stars lying to the {\em red} of the instability strip (Red Clump stars) must be
(comparatively) young  or  metal-rich, or both (see Monaco et al. \cite{lbhb},
and B06 for the  interpretation of the HB  morphology of Sgr in this context). 
Monaco et al.(\cite{lbhb})  noted that both the BHB and RC sequences observed 
at the center of Sgr have a clear counterpart in the Color-Magnitude Diagrams 
(CMDs) of Stream fields obtained by N02 using data from the Early Data Release 
of the Sloan Digital Sky Survey (SDSS).  
Here we perform a direct comparison of the relative abundance of BHB and RC 
stars in the Sgr core and in one of the Stream fields studied by N02,
to search for a population gradient. While the results we obtain are
qualitatively similar to those of C06, they refer to a different and more
distant portion of the Stream, and provide an
on-field demonstration that the whole Sgr Stream can be studied with this
technique.

\subsection{The data}

To obtain a representative sample of the stellar population in the core of Sgr
we take the photometry of the $1\degr \times 1\degr$ wide field located $\sim
2\degr$ eastward of the galaxy center, recently presented by B06, i.e. the
Sgr34 field [(l,b)  $\simeq(6.5\degr,-16.5\degr$), see B06 for details]. N02
provided two very clear detections of the Sgr Stream in their fields
S341+57-22.5 and S167-54-21.5, where the first two numbers of the names are the
mean Galactic longitude and latitude. While the BHB is clearly
visible in the CMD of S167-54-21.5, the signal of the Stream RC is too weak in
this field to derive fully reliable star counts, due to various intrinsic
factors that will be discussed elsewhere (Bellazzini et al., in preparation).
Hence exclude this field from the analysis and we focus on S341+57-22.5  (S341
hereafter, for brevity), a $\sim 62.5$ deg$^2$ strip enclosed within $200\degr \le
RA\le 225\degr$ and $-1.25\degr\le Dec\le 1.25\degr$, that samples a portion of
the leading arm of the Stream, lying more than 40 kpc away from us in the
Northern Galactic hemisphere (N02, Law et al. \cite{law}). 
The g$^*$,r$^*$ photometry by N02 has been transformed into Jonhsons-Cousins
B,V according to Smith et al. (\cite{smith}). 
The photometry has been corrected for extinction following B06. 
Note that the average reddening in the S341 field is $E(B-V)=0.04$ mag.

\section{HB star counts} 

In Fig. \ref{HBSEL} we present reddening-corrected B, B-V CMDs focused on the
HB of fields within the main body of Sgr and of the S341 field in the Sgr
Stream. We chose the B, B-V CMD because in this plane a larger section of the 
BHB sequence appears nearly horizontal, with respect to V, B-V diagrams.
This will allow a better background subtraction, according to the method that
will be described below. Conversely, the RC, that appears tilted in
the CMDs of Fig.~\ref{HBSEL}, is nearly perfectly horizontal in V, B-V diagrams.

The filled circles superposed in the CMD of SGR34 are the ab type RR Lyrae
variables we counter-identified in a field located $\sim 2^o$ westward  of the 
center of Sgr (SGRW) from the catalogue of Sgr variables by 
Cseresnjes (\cite{cser}). The wide area surveyed by Cseresnjes (\cite{cser})
does not include the SGR34 field but has a large overlap with SGRW.
The photometry is from the same survey of the SGR34 
field, the data reduction has been performed as in B06.
In our dataset
the variables are observed at random phase: their color-magnitude distribution
is useful here to indicate the position of the instability strip in our
diagrams, that will help in the selection of relatively {\em pure} BHB and 
RC samples. 
The RC-labeled color strip $0.7\le (B-V)_0\le 0.95$
encloses the Red Clump population in the main body (at $B_0\sim 18.7$, upper and
middle panel of Fig.~\ref{HBSEL}) and in
the S341 field (at $B_0\sim 20.0$, lower panel; the fainter mean magnitude 
with respect to
SGR34 is due to a difference in distance between the populations sampled by the
two fields, see N02 and Monaco et al. \cite{ltip}). 
The BHB-labeled color strip
($0.0\le (B-V)_0\le 0.2$), on the other hand, encloses the more populated portion
of the BHB, far to the blue of the ab RR Lyrae, at $B_0\sim 18.0$ in
SGR34 and at $B_0\sim 19.3$ in S341. 
Obviously there are BHB stars bluer than this limit; the adopted selection is
optimally suited for the background subtraction technique described below.
Since the comparison is purely differential this does not affect the results
discussed below, but the adopted selection
does not allow a complete census of BHB stars\footnote{This is why we
find a smaller fraction of BHB stars relative to the sum of BHB and RC stars
here than in Monaco  et al. (\cite{lbhb}). In that paper a more generous
selection box was adopted, including also the vertical part of the BHB.}.
It can be seen from Fig.~\ref{HBSEL} that the RC in S341 is slightly bluer than
in SGR34 (by $\simeq 0.05$ mag). This may be interpreted as a signature
of a difference in mean metallicity between the two populations. However, it
cannot be excluded that  relatively small color differences may be produced by
less-than-perfect transformations between the $g^*,r^*$ and $B,V$, and/or by
errors in reddening or calibration, while star counts should be essentially
unaffected by these problems (see below). 

   \begin{figure}
   \centering
   \includegraphics[width=7.0cm]{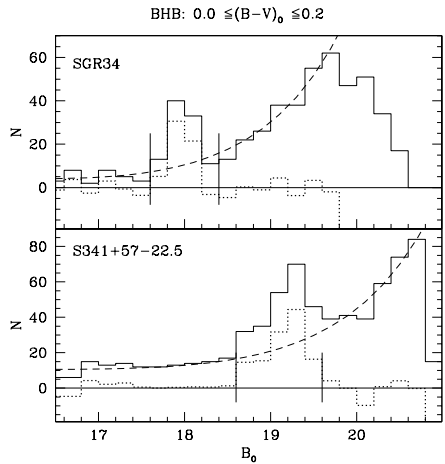}
   \caption{LFs of the BHB color strip. The continuous-line histograms are
   the observed LFs, the long-dashed lines are the best-fit curves to the
   underlying ``background'' LF outside the BHB peaks, 
   the dotted-line histograms
   are the residual of the $observed~LF~-~fitted~LF$ subtraction.
   The thin vertical segments enclose the range in which we count BHB stars  (the
   residuals of the fit) to
   obtain $N_{BHB}$.}
              \label{conBHB}
    \end{figure}
%

   \begin{figure}
   \centering
   \includegraphics[width=7.0cm]{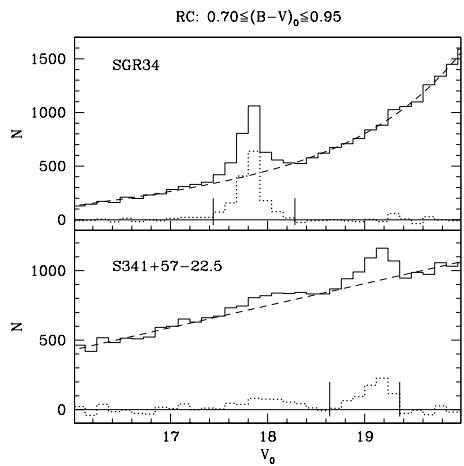}
   \caption{LFs of the RC color strip. The symbols are the same as in Fig.~2.
   The thin vertical segments enclose the range in which we count RC stars.
}
              \label{conRC}
    \end{figure}

While clearly visible in all the presented CMDs, both the RC and the BHB
features are seriously affected by contamination from unrelated sources, either
Sgr stars in different evolutionary phases\footnote{For example, most of the
faintest stars in the Fig.~1 CMD within the BHB color strip belong to the Blue Plume
population of Sgr (see B06).} or foreground stars
of our Galaxy. To remove the contribution of this generic ``background signal''
from star-counts we proceeded as follows: (a) we construct the Luminosity
Function (LF) of the stars lying in a given color-strip (RC or BHB), 
(b) we fit the LF outside the peak associated with the RC and/or BHB features,
(c) we subtract the fitted LF to the observed one, and (d) we sum the
residuals within the narrow region enclosing the RC or BHB peak.
A similar approach has been used before, see,
for example, Harbeck et al. (\cite{harb}).
We adopt $B_0$ LFs for the BHB strip and $V_0$ LFs for the RC strip to obtain
the cleanest signal as possible for the two features that greatly differ in
color (temperature).
The observed LFs are plotted in Fig.~\ref{conBHB} for BHB stars,
and in Fig.\ref{conRC} for the RC stars. RC and BHB peaks are clearly visible in
all the presented LFs, hence the adopted technique can be reliably
used. The fact that the BHB and RC peaks in the S341 LFs are somewhat
wider than those in SGR34 is probably due to an intrinsic difference in the
distance profile along the line of sight of the two considered populations, the
core of Sgr being more compact than the considered portion of the Stream. 
The results of the background-subtracted star counts are summarized in
Table~\ref{tab1}. 
The square root of the total number of stars counted
in the reported magnitude windows (i.e. before subtraction of the 
background), are adopted as errors on the star counts ($N_{BHB}$, $N_{RC}$), 
according to Poisson's statistic.
To check the effects of color shifts (due to
metallicity gradients and/or errors in reddening or in the photometric
calibration) we repeated the whole procedure after shifting
the selection strips by $\pm 0.03$ mag. The resulting $N_{BHB}/N_{RC}$ ratios 
are unchanged, within the uncertainty, with respect to the values reported
in Table~1. We also checked the case of a shift of -0.05 mag applied only to the
RC strip of S341, to take into account the observed color shift of the Stream
RGB in this field, and also in this case the change of $N_{BHB}/N_{RC}$ is
smaller that the derived uncertainty. 

Table~1 shows that the ratio of the number of BHB stars to the
number of RC stars ($N_{BHB}/N_{RC}$) is more than five times larger in the 
portion of the Stream sampled by the S341 (at $D_{\sun}\simeq 45$ kpc from us,
and more than 50 kpc away from the center of the parent galaxy) than within the
core of the Sgr galaxy (at $D_{\sun}\simeq 26$ kpc from us). The detected
difference between the two fields is significant at the $\simgt 4.8\sigma$ 
level; 
given the hypothesis that the HB morphology in S341 is the same as in SGR34, 
we expect $N_{BHB}=24\pm 5$ in the Stream field, while we observe 
$N_{BHB}=122\pm 15$. 
Hence, BHB stars are much more abundant in the Stream 
(15\% of the whole RC+BHB population) than in the main body of Sgr 
(just 3\%, with the adopted selection; see footnote 1). 
This is a strong indication that old and metal-poor stars are much
more represented in the population that was stripped from Sgr in the past than
in the presently bound core, providing an independent confirmation and extension
of the results by M03 and C06.

The techniques adopted by M03 and C06
may be limited in their application by a low sensitivity (as the infrared colors
of M-giants, M03) or by the faintness of targets in distant regions of the
Stream (as the high-resolution spectroscopy of individual stars by C06). 
The approach adopted here cannot provide a direct probe of the metallicity
distribution as done by C06, nevertheless it
relies on a very sensitive diagnostic (i.e.
the HB morphology) that can be successfully measured out to very large
distances. 
Hence, it may give the opportunity to trace population gradients along the
whole extension of the Sgr Stream. Note that the procedure described here and
the data reported in Table~\ref{tab1} provide the basis (in particular the ``Sgr
core zero-point'' of $N_{RC}/N_{BHB}$) to measure the population
gradient in an homogeneous scale to anyone having access to CMDs of the Stream
deep enough to sample the whole HB. 
This may even allow one to probe differences in
stellar populations between wraps of the Stream lying in the same direction but
at different distances, such as those seen by Belokurov et al. (\cite{belok}) 
and modeled by Fellhauer et al. (\cite{fel}).

It is apparent from Fig.~2 and Fig.~3 that the accurate location (in
magnitude) of features like the RC and the BHB may provide powerful constraints
on the distance and the structure of the Stream. We will discuss in detail these
applications in a forthcoming contribution (Bellazzini et al., in
preparation).

   \begin{table}
      \caption[]{Star counts in the BHB and RC}
         \label{tab1}
\centering                              
\begin{tabular}{l c c}        
\hline\hline                            
          &   Sgr34      & S341+57-22.5 \\    
\hline                                  
$N_{BHB}$ & $54\pm 10$    & $122 \pm 15$ \\
$N_{RC}$  & $1542\pm 67$ & $686 \pm 78$ \\
&&\\
${N_{BHB}\over{N_{RC}}}$& $0.035 \pm 0.007$ & $0.18 \pm 0.03$\\
&&\\
${N_{BHB}\over{N_{RC}+N_{BHB}}}$& $0.03$ & $0.15$\\
\hline                                                
\end{tabular}
\end{table}

\section{Conclusions}

We have detected a $\simgt 4.8\sigma$ difference in the ratio of BHB to RC stars
between the core of Sgr and a distant portion of its tidal Stream. BHB stars are
5 times more abundant - relative to RC stars - in the considered Stream field 
than within the main body of the galaxy, a clear signature of the presence of
an age/metallicity gradient along the Sgr remnant.

C06 clearly states that the observed metallicity gradient within the Stream
cannot be caused by ``an intrinsic variation of the instantaneous mean
metallicity of Sgr with time'', because the chemical enrichment of Sgr was
essentially completed much before than the Stream stars that display the
gradient effect were stripped from the main body of their parent galaxy (see
also B06). This leaves the presence of a radial metallicity/age gradient in the
progenitor as the only viable explanation for the observed core/Stream gradient
 (note, however, that the simultaneous presence
of stars stripped from Sgr in different peri-Galactic passages in the same
portion of the Stream may also play a role; see Fellhauer et al. 2006). 
C06 notes also that the implied radial
gradient should have been significantly stronger than that typically encountered
in dwarf spheroidal galaxies. 
A comparison
with the results by Harbeck et al. (\cite{harb}) shows that in the galaxies of
their sample the $N_{BHB}/N_{RC}$ ratio typically varies by a factor of $<3$,
over the considered radial range, that is less than the factor of 5 we observe
between SGR34 and S341. 
However changes of a factor of $\simgt 8$ are observed
in Sculptor and Sextans, hence the value found here does not appear
exceptional, although the exclusion of the hottest stars from our BHB selection
may hide a larger difference. 

According to the N-body models by Law et al. (\cite{law}) the stars populating
the branch of the Stream sampled by the S341 field have been stripped during
different orbital revolutions, in the past. Independently of the
adopted model (adopting a {\em spherical}, {\em oblate} or {\em prolate}
Milky Way halo), only $\sim 30$\% of the stars in this part of the Stream have
been lost from the main body during the current peri-Galactic passage or 
during the immediately previous one: hence the observed population 
should be dominated by stars lost more than $\simeq 1.5-2$ Gyr ago (see also
Belokurov et al. \cite{belok} and Fellhauer et al. \cite{fel}). Thus the
observed HB morphology should reflect the stellar mix in the outer regions of
the Sgr progenitor as it was 2 to 4 peri-Galactic passages ago. We
conclude that such a stellar mix was significantly more
metal-poor, on average, than that observed in the present-day bound remnant.

\begin{acknowledgements}
This research is supported by the INAF-PRIN2005 grant CRA 1.06.08.02.
H.J.N. acknowledge the support of NSF through the grant AST03-07571.
The N-body models by Law et al. (2005) have been retrieved from the 
homepage of S. Majewski ({\tt http://www.astro.virginia.edu/~srm4n/Sgr/}).
We are grateful to the referee (A. Aparicio) for his useful suggestions and
comments.

\end{acknowledgements}

\end{document}